\documentclass[conference]{IEEEtran}
\IEEEoverridecommandlockouts
\usepackage{cite}
\usepackage{amsmath,amssymb,amsfonts}
\usepackage{algorithmic}
\usepackage{graphicx}
\usepackage{textcomp}
\usepackage{xcolor}
\usepackage{tabularx,booktabs}
\def\BibTeX{{\rm B\kern-.05em{\sc i\kern-.025em b}\kern-.08em
    T\kern-.1667em\lower.7ex\hbox{E}\kern-.125emX}}
\begin{document}

\title{Pseudo Supervised Solar Panel Mapping based on Deep Convolutional Networks with Label Correction Strategy in Aerial Images\\

}

\author{
 \IEEEauthorblockN{Jue Zhang\IEEEauthorrefmark{1},
  Xiuping Jia\IEEEauthorrefmark{2}, 
  Jiankun Hu}
 \IEEEauthorblockA{School of Engineering and Information Technology\\
  The University of New South Wales\\
  Canberra, ACT 2600, Australia\\
  Email: \IEEEauthorrefmark{1}jue.zhang@student.unsw.edu.au,
  \IEEEauthorrefmark{2}x.jia@adfa.edu.au}}

\maketitle

\begin{abstract}
Solar panel mapping has gained a rising interest in renewable energy field with the aid of remote sensing imagery. Significant previous work is based on fully supervised learning with classical classifiers or convolutional neural networks (CNNs), which often require manual annotations of pixel-wise ground-truth to provide accurate supervision. Weakly supervised methods can accept image-wise annotations which can help reduce the cost for pixel-level labelling. Inevitable performance gap, however, exists between weakly and fully supervised methods in mapping accuracy. To address this problem, we propose a pseudo supervised deep convolutional network with label correction strategy (PS-CNNLC) for solar panels mapping. It combines the benefits of both weak and strong supervision to provide accurate solar panel extraction. First, a convolutional neural network is trained with positive and negative samples with image-level labels. It is then used to automatically identify more positive samples from randomly selected unlabeled images. The feature maps of the positive samples are further processed by gradient-weighted class activation mapping to generate initial mapping results, which are taken as initial pseudo labels as they are generally coarse and incomplete. A progressive label correction strategy is designed to refine the initial pseudo labels and train an end-to-end target mapping network iteratively, thereby improving the model reliability. Comprehensive evaluations and ablation study conducted validate the superiority of the proposed PS-CNNLC.
\end{abstract}

\begin{figure*}[h]
    \centering
    \includegraphics[width=18cm]{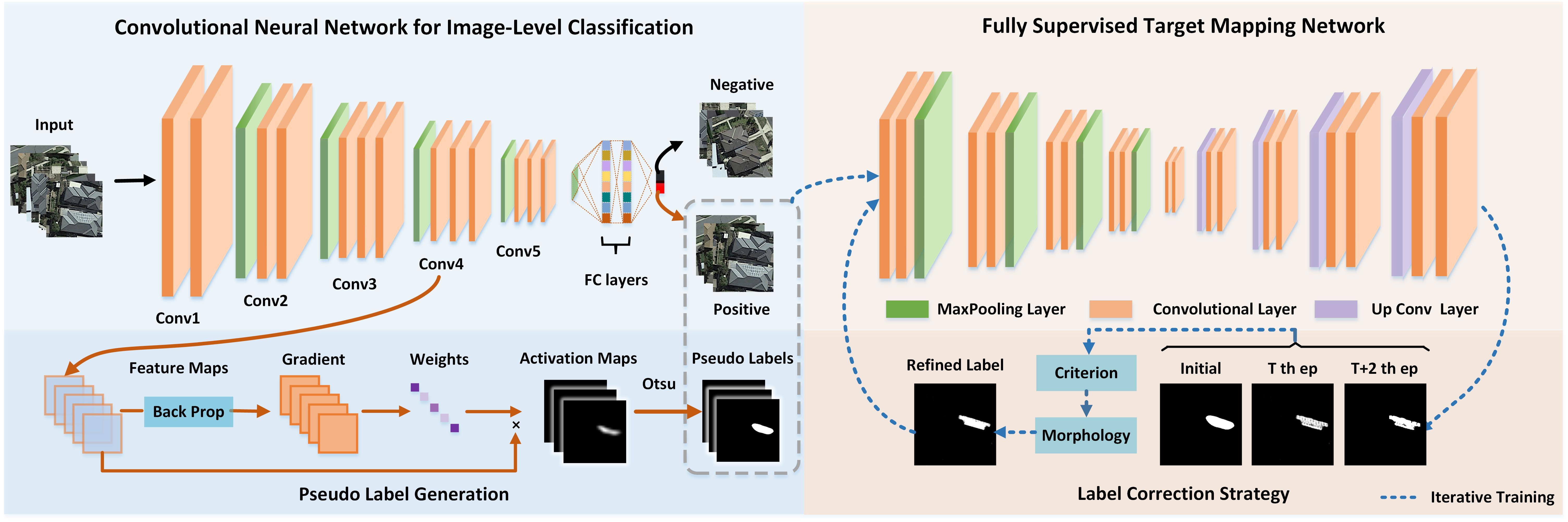}
    \caption{The flowchart of the proposed pseudo supervised solar panel mapping method.}
    \label{fig:my_label}
\end{figure*}
\section{Introduction}
Solar energy has been the fastest-growing type of renewable energy worldwide in recent years \cite{panwar2011role, haegel2017terawatt,alam2013approach}. The recent improvement of solar photovoltaic (PV) makes the small-scale or distributed solar such as rooftop-PV possible. However, the intermittency nature and sparse distribution of small-scale solar pose challenges to grid compatibility [4, 5] and operation. Monitoring solar panel installations including their locations and sizes is increasingly meaningful to government and electricity companies for planning power grids and minimising potential risks. Thanks to the improvements of remote sensing techniques, very high-spatial-resolution remote sensing images are now available for automatic and regular inspection of solar panel distributions at a large scale.   

Automatic solar panel mapping in remote sensing images has been attempted in recent years with machine learning techniques. They can be divided into two branches: unsupervised methods and supervised methods. Unsupervised methods often use template matching \cite{wang2018photovoltaic}. To date, supervised methods is the mainstream in solar panel mapping as they have superior capability in distinguishing small scale objects in complex scenes. Supervised object extraction used in \cite{malof2015automatic,malof2016automatic} for solar panel mapping is composed of three main steps: initial localization, feature extraction by handcrafted descriptors, and decision-making via machine learning algorithms. For feature extraction, local color statistic (LCS) feature, texture feature and shape feature are frequently investigated, and support vector machine (SVM) \cite{malof2015automatic,lisolarfinder} and random forest (RF) \cite{malof2016automatic} classifiers are popular choices in the classification phase. Some supervised methods regard the task as a spectral un-mixing problem \cite{karoui2018detection,karoui2019partial} with the aid of hyperspectral remote sensing data.

With the development of deep learning techniques \cite{schmidhuber2015deep,lecun2015deep}, convolutional neural networks (CNNs) have also been applied in the last a few years to solar panel mapping \cite{yuan2016large,castello2019deep,malof2017deep,yu2018deepsolar,zhuang2020automatic}. Yuan et al. proposed a large scale solar panel detection method based on deep convolutional networks in a fully supervised manner, in which feature maps generated by different layers are finally stacked and then fed into a single convolutional layer to produce a dense prediction \cite{yuan2016large}. Later on, Malof et al. \cite{malof2017deep} used a network similar to the VGGNet \cite{simonyan2014very} and investigated the impacts of different training strategies. Different from the previous work, Wang et al. considered the detection of PV arrays as a semantic segmentation task, and introduced SegNet \cite{badrinarayanan2017segnet} to produce dense predictions of PV arrays’ locations and sizes. U-net \cite{ronneberger2015u} and its improved versions such as cross learning-driven U-net method (CrossNets) \cite{zhuang2020automatic} were also introduced to identify the solar installations.

For fully supervised CNN-based methods, manually annotating pixel-wise ground-truths or bounding boxes for providing a large number of training samples is a challenging task as it is usually human-force intensive \cite{zhou2018brief}. In contrast,  weaker forms of annotations, for instance image-wise labels indicating whether there are targets in the image or not, are efficient and easy to obtain. This is particularly the case in the field of remote sensing, where the generation of accurate pixel-wise labels relies heavily on expert knowledge. Hence, weakly supervised learning, which uses these coarse annotations to supervise the training of deep learning models is becoming increasingly meaningful \cite{bency2016weakly,choe2019attention,zhu2019learning}. Class activation mapping (CAM) \cite{zhou2016learning} and gradient-weighted class activation mapping (GradCAM) \cite{GRADCAM} and deep feature maps \cite{bency2016weakly} are works in weakly supervised object localization. A number of weakly supervised frameworks are also proposed for object detection and segmentation in remote sensing images \cite{han2014object,HWSL,ma2019weakly,li2018deep}. Yu et al. developed the first weakly supervised deep learning framework for solar panel mapping in very high resolution aerial images, named ‘DeepSolar’ \cite{yu2018deepsolar}, with Inception v3 classification network \cite{szegedy2016rethinking} and CAM. However, as deep representations learned for CNNs only correspond to the most discriminative part of the objects, these methods may fail to provide a complete object content. Although a number of post-processing strategies including dense conditional random field (dense CRF) and Markov random field (MRF) have been proposed to refine the results, difficulties remain in providing high detection accuracy and reducing computational complexity. 

In this paper, we propose a pseudo supervised solar panel mapping method based on deep convolutional networks with label correction strategy, aiming to develop an end-to-end network for accurate and efficient solar panel extraction with only image-wise annotations. The main contributions of the proposed method are two-fold:

(1) We construct an end-to-end fully supervised target mapping network by taking advantages of pseudo labels generated by weakly supervised learning, thereby reducing the computational complexity in target mapping.

(2) During the training course, a progressive label correction strategy based on refinement criteria and morphological filtering is proposed to refine the initial pseudo labels. The target mapping network is trained iteratively with progressively corrected labels to gain further improvement in the model development. 
\section{Methodology}
 The proposed pseudo supervised solar panel mapping method contains three parts:  convolutional neural network for image-level classification, pseudo label generation based on gradient-weighted class activation mapping, and fully supervised target mapping network with label correction strategy. The flowchart of the proposed method is shown in Figure 1.
\subsection{Convolutional Neural Network for Image-level Classification}
For the classification network, we adopt a network architecture similar to VGG16 \cite{simonyan2014very}, proposed by Simonyan et al., which contains 13 convolutional layers and three fully connected layers. All the convolutional layers have a very small $3\times 3$ receptive field, fixed stride (1 pixel), and 1 pixel padding, with a rectified linear unit ($Relu$) added to every layer to increase non-linearity. As solar panels are relatively small objects in aerial images, the employment of small receptive fields can help preserve the boundaries and improve completeness.  Maxpooling with stride 2 is carried out five times over a  $2\times 2$ window, dividing the stacked convolutional layers into five blocks with different sizes of feature maps, which are denoted as $Conv1$, $Conv2$, $Conv3$, $Conv4$, $Conv5$. In the following context, we use $Conv n\_p$ to denote the $p^{th}$ convolutional layers in $n^{th}$ block in the classification network.  

In our work, the provided image-wise annotations indicate whether or not there are solar panels in the image (positive or negative samples), then the classification network is trained to solve a binary classifications problem. Two fully connected (fc) layers with 256 channels and another one with 2 channels are added before the soft-max layer. The 2-dimension output of the classification network is the probabilities of the input sample belonging to positive or negative cases, respectively. The cross entropy loss is utilized to train the classification network with all the convolutional layers' weights pre-trained with the ImageNet datasets \cite{russakovsky2015imagenet}. 

\subsection{Pseudo Label Generation}
As we attempt to construct a fully supervised model for solar panels mapping, initial solar panel extraction can be obtained via target localization from the positive samples identified by the well-trained image-level classification network. CAM provides a good idea to localize the regions having decisive impacts on the classification scores. It maps the predicted probabilities of different classes back to the feature maps yielded by previous convolutional layers to generate informative maps, in which the most discriminative regions (Solar panels) in the image will be highlighted. The activation map for class $c$ can be computed as follows:
\begin{equation}
    f\left ( I ;\Theta \right )_{c} =\sum_{k}^{}w_{c}^{k}\frac{1}{N}\sum_{x,y}^{}F^{k}(x,y),
\end{equation}
\begin{equation}
    CAM_{c}=\sum_{k}^{}w_{c}^{k}F^{k}(x,y),
\end{equation}
where the input image with the size of $h\times w$ is denoted with $I\in \mathbb{R}^{h\times w}$. $f\left ( \cdot ;\Theta \right )$ is the nonlinear mapping function represented by the classification network with a parameter set $\Theta$. $f\left ( I ;\Theta \right )_{c}$ is the predicted score of image $I$ for class $c$. $F^{k}(x,y)$ is the $k^{th}$ channel of feature map $F$, and $w_{k}^{c}$ is the weight of the $c^{th}$ neuron in the fully connected layer. $\frac{1}{N}\sum_{x,y}^{}(\cdot )$ is the global average pooling (GAP) operation, $N=h\times w$. $(x,y)$ denotes the pixel location. 

CAM, however, has obvious drawbacks: it requires GAP to transfer every channel of the feature maps into a neuron, which may bring negative impacts to classification accuracy \cite{GRADCAM}. To solve this issue, Selvaraju et al. \cite{GRADCAM} proposed a gradient-weighted class activation mapping with the gradient flowing in the network, and the GAP operation is on longer needed in the network architecture.

Specifically, let $F_{i}^{k}$ indicate the $k^{th}$ channel in the feature maps produced by $i^{th}$ convolutional layers. Instead of computing weights via GAP and fully connected layers, GradCAM calculates the gradient of class score $f\left ( I ;\Theta \right )_{c}$  to $F_{i}^{k}$. The final activation map is obtained by combine feature maps with the global average pooled gradients $\widehat{w}_{c}^{k}$. The specific calculation is shown as follows: 
\begin{equation}
\widehat{w}_{i,c}^{k}=\frac{1}{N_{i}}\sum_{x,y}^{}\frac{\partial f\left ( I ;\Theta \right )_{c}   }{\partial F_{i}^{k}},
\end{equation}
\begin{equation}
GradCAM_{i, c} =Relu\left \{ \sum_{k=1}^{}\widehat{w}_{i,c}^{k} F_{i}^{k}(x,y) \right \}.
\end{equation}

We compare the activation maps generated by convolutional layers in different depth and have observed that deeper convolutional layers provide coarser localization but better coverage, while activation maps obtained with shallow convolutional layers only highlight small part of the discriminative regions with better boundary preservation, as shown in Figure 2. 
\begin{figure}[htbp]
    \centering
    \includegraphics[width=8.5cm]{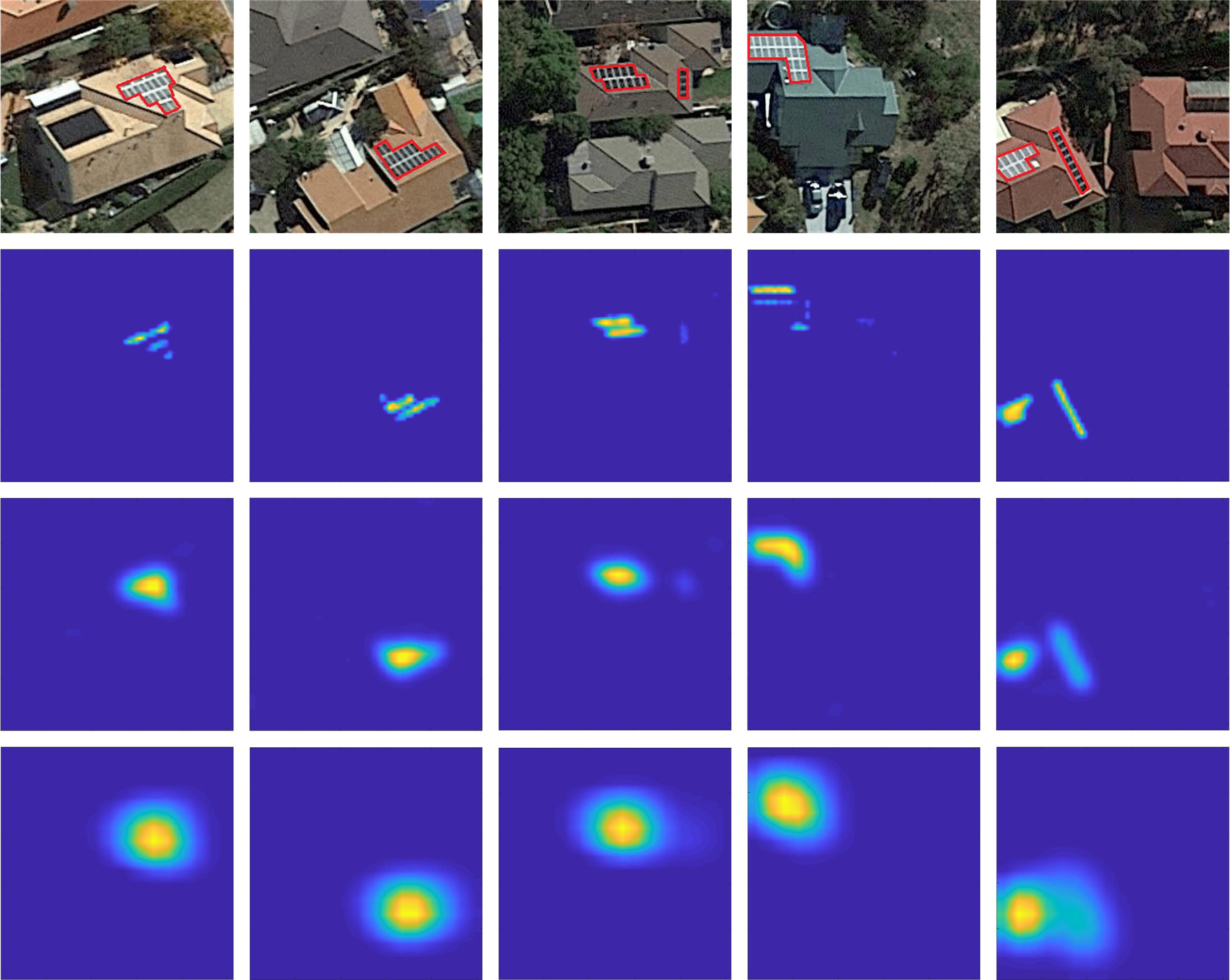}
    \caption{Activation maps with feature maps produced by different convolutional layers. First row: original images with solar panels highlighted by red lines. Second to Forth row: activation maps generated with $conv3\_3$, $conv4\_3$ and $conv5\_3$ layer. Yellow pixels indicate high activation values while blue ones indicate low values.}
    \label{fig:my_label}
\end{figure}
Therefore, we select feature maps produced by $conv4\_3$ to generate activation maps. Then we propose to use Otsu method \cite{otsu1979threshold} to binarize the activation maps to finalize the target labels.

As the original training images are limited, unlabeled images are then utilized, from which possible positive images can be recognized with the trained classification network. GradCAM and Otsu thresholding are then applied to generate more target labels. They are named as the initial pseudo labels $GT^{0}$, as they are self-generated from the classification network. The pseudo labels of the original training samples are also included in $GT^{0}$ for simplicity. 
\subsection{Fully Supervised Target Mapping Network with Label Correction Strategy}
\subsubsection{Fully Supervised Target Mapping Network}
In this part, we train a fully supervised target mapping network with pseudo labels using the method presented in Section A and B. Although we add extra training samples by considering unlabeled data by the trained classification network, the number of training samples is limited. Considering this, we construct the network in an encoder-decoder architecture similar to the U-net \cite{ronneberger2015u}.
In the encoder path, four encoder blocks are added, with each containing a  $2\times 2$ maxpooling layer and two $3\times 3$ convolutional layers with stride 1. With the path goes deeper, the channels of feature maps doubles after every maxpooling operation.  $E_{i}^{2k}$ denotes the output of the $i^{th}$ encoder block with $2k$ channels, $i=1,2,3,4$. $Conv_{3}$ denotes the convolutional layer with $3\times 3$ kernel and 1 pixel stride: 
\begin{equation}
E_{0} = I,
\end{equation}
\begin{equation}
E_{i}^{2k}=\downarrow Conv_{3}(Conv_{3}(E_{i-1}^{k})),
\end{equation}
where $I\in \mathbb{R}^{h\times w}$ is the input image with the size of $h\times w$ . $\downarrow $ denotes the maxpooling operation.  $E_{i-1}^{k}$ denotes the output of the $(i-1)^{th}$ encoder block with $k$ channels.
\begin{figure}[htbp]
    \centering
    \includegraphics[width=8.5cm]{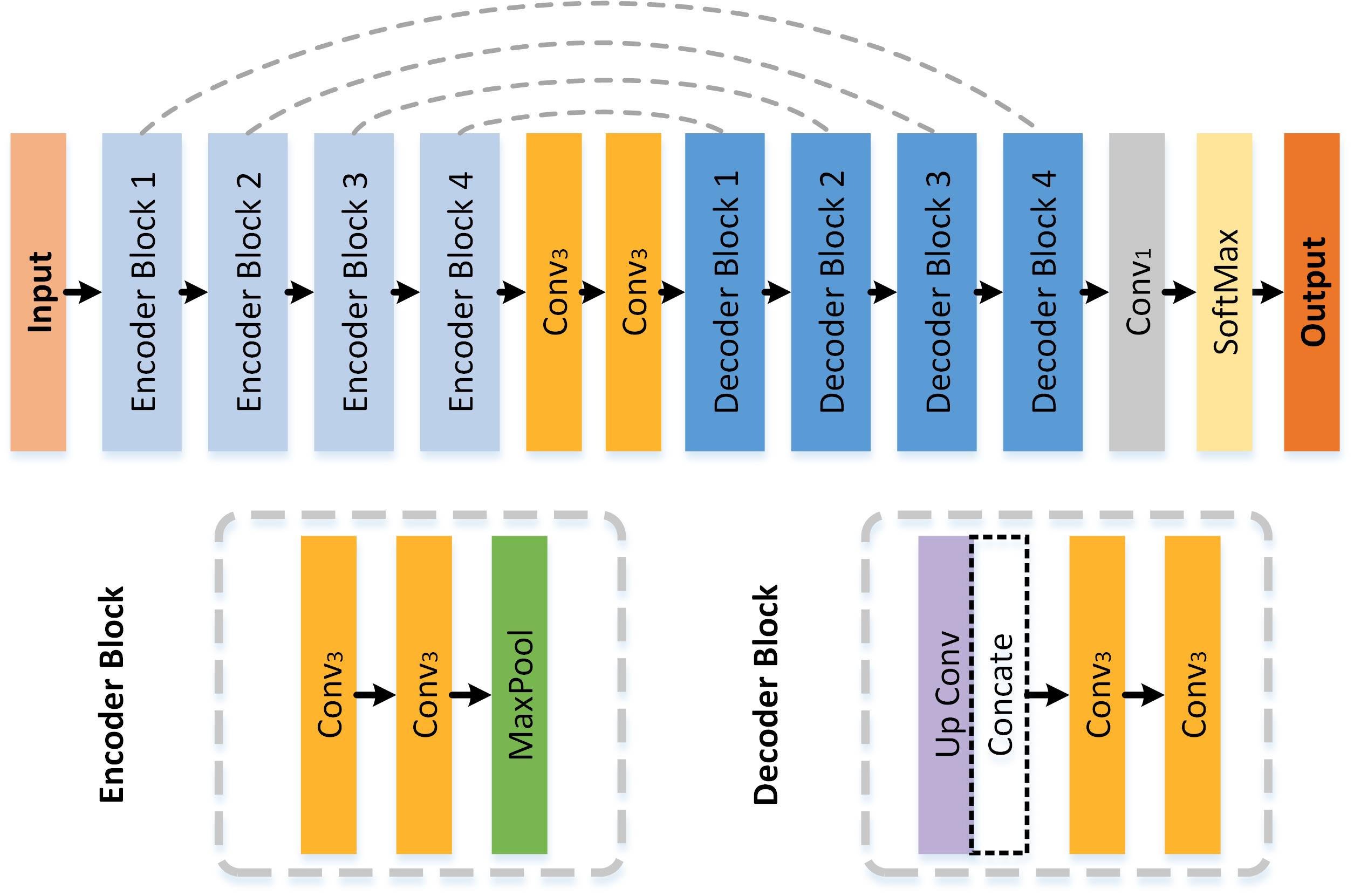}
    \caption{The architecture of the fully supervised mapping network.}
    \label{fig:my_label}
\end{figure}
Then, two convolutional layers with $3\times 3$ kernel and 1 pixel stride are added to increase the channels of feature maps. $E_{4}^{512}$ is the output of the last encoder block and $B^{1024}$ is the output of the two convolutional layers:
\begin{equation}
B^{1024} =Conv_{3}(Conv_{3}(E_{4}^{512})).
\end{equation}

In the decoder path, every decoder block comprises a  $3\times 3$ transpose convolutional layer (up conv layer) with 2-pixel stride and two $3\times 3$ convolutional layers with 1-pixel stride. In order to combine low-level and high-level features, skip connections are added to every corresponding block in the encoder and decoder path in a concatenation way. The input of the first decoder block is $B^{1024}$ and the feature maps $G_{4}^{512}$ generated by the second convolutional layers in the forth encoder block. $D_{1}^{512}$ is the output of the first decoder block:
\begin{equation}
D_{1}^{512}=Conv_{3}(Conv_{3}([\uparrow B^{1024},G_{4}^{512}])).
\end{equation}

The input of the $(j+1)^{th}$ decoder block is the concatenation of the output of the previous block $D_{j}^{k}$ in the decoder path and the feature maps generated by the second convolutional layers in corresponding block $G_{4-j}^\frac{k}{2}$ in the encoder path, $j=1,2,3$:
\begin{equation}
D_{j+1}^{\frac{k}{2}}=Conv_{3}(Conv_{3}([\uparrow D_{j}^{k},G_{4-j}^\frac{k}{2}])),
\end{equation}
where $\uparrow $ denotes the transpose convolutional layer.

Finally, $1\times 1$ convolutional layer and a soft-max layer are added to generate a 2-channel output, which has the same size with the input image. $D_{4}^{64}$ is the output of the decoder path:
\begin{equation}
    f\left ( I ;\Phi  \right ) =Softmax(Conv_{1}(D_{4}^{64})).
\end{equation}
where $f\left ( I ;\Phi  \right )$ is the output of the mapping network with a parameter set $\Phi$.

As solar panels is the minor class compared to other major urban classes in the background of aerial images, we use weighted cross entropy instead of cross entropy to train the network:    
\begin{equation}
\alpha =  \frac{N-N_{fore}}{N},
\end{equation}
\begin{equation}
Loss = - \sum_{x,y}^{}(\alpha \cdot GT \cdot log(f_{0})+(1-\alpha )\cdot (1-GT)\cdot log(f_{1})),
\end{equation}
where $N_{fore}$ is the number of pixels belonging to the foreground, i.e. solar panels in the ground-truth $GT$. $f_{0}$ and $f_{1}$  are the first and second channels in the final output, respectively.
\begin{figure*}[htp]
    \centering
    \includegraphics[width=18cm]{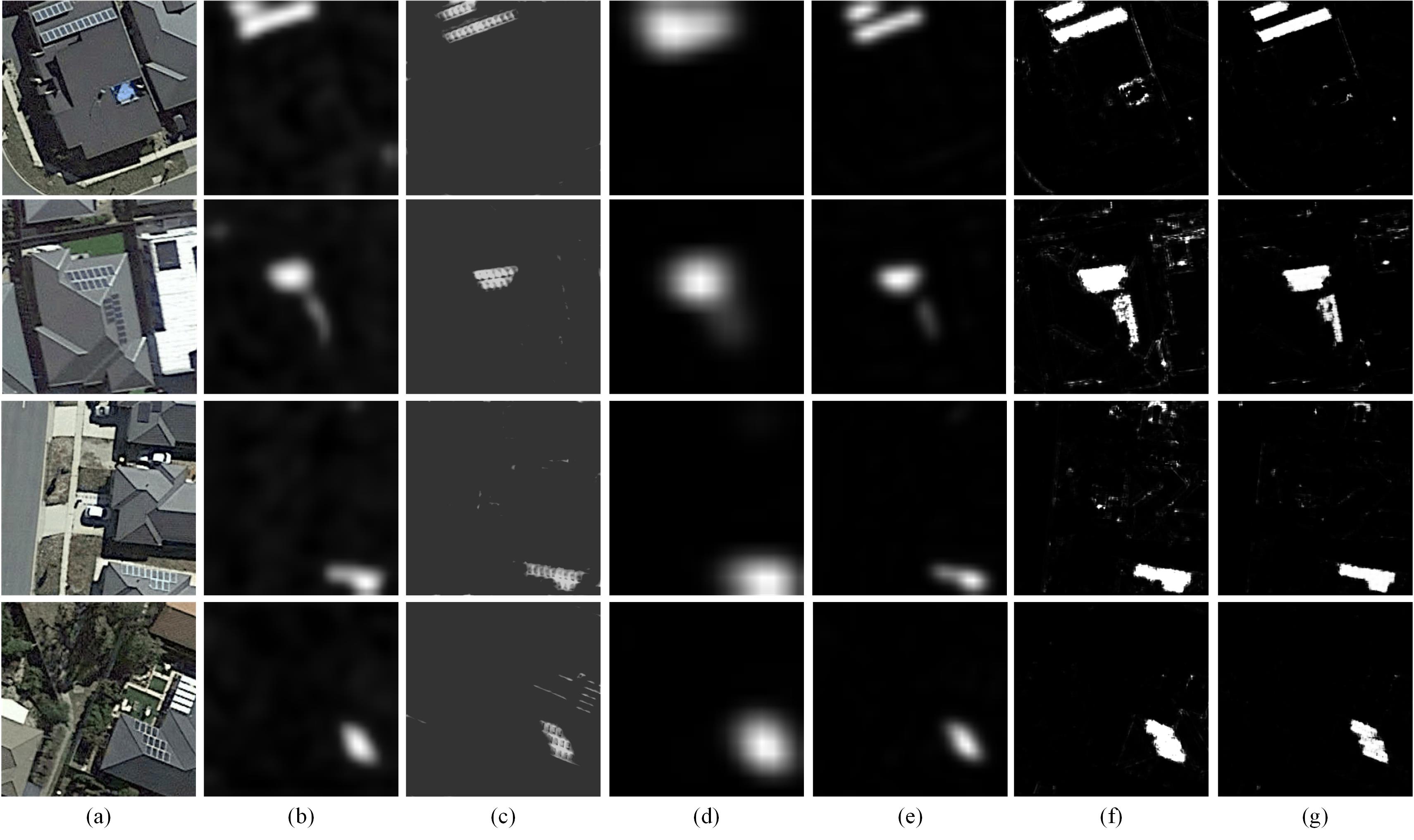}
    \caption{Solar panel mapping results by different methods. (a) Original images. (b) DeepSolar. (c) HWSL. (d) GradCAM$_{5}$. (e) GradCAM$_{4}$. (f) PS-CNN. (g) PS-CNNLC.}
    \label{fig:my_label}
\end{figure*}
\subsubsection{Label Correction Strategy}
With the uncertainty in the coarse labels provided, the fully supervised network can hardly be trained properly. For example, some rooftop pixels may be identified as solar panels in the initial pseudo labels, which would lead to many false positives; some small solar panels may be omitted, which would lead to many false negatives. In order to address these problems, we propose a progressive label correction strategy to refine the initial pseudo labels and train the mapping network iteratively. 

Firstly, we design three criteria for evaluating the output of the mapping network. Given the output $f^{T}$ after $T^{th}$ epochs training, the output $f^{T+2}$ after $(T+2)^{th}$ epochs training is examined based on the following expectations:
\begin{itemize}
\item  The size of extracted regions in produced output should be within the general sizes of residential solar panels in aerial images. Based on our observation, the minimum number of panels installed on a rooftop is rarely fewer than 5, and the maximum is often not over 60. 
\item Although the initial pseudo label provides localization with certain noise, the output should not be totally different from the initial ones. It is, however, noteworthy that some small panels may be omitted in the initial labels. 
\item The area of extracted regions should not decrease or increase significantly in the two successive training epochs. A dramatic change indicates unstable prediction.
\end{itemize}

The proposed rules can be summarized as follows:
\begin{equation}
f^{T+2}_{fore} \in (\beta_{1}\cdot N,\beta_{2}\cdot N),
\end{equation}
\begin{equation}
f^{T+2}_{fore} \in (\gamma _{1}\cdot GT^{0}_{fore},\gamma _{2}\cdot GT^{0}_{fore}),
\end{equation}
\begin{equation}
f^{T+2}_{fore} \in (\delta  _{1}\cdot f^{T}_{fore},\delta  _{2}\cdot f^{T}_{fore}).
\end{equation}
where $f^{T+2}_{fore}$ denotes the number of pixels belonging to the fore-ground, i.e. solar panels after $f^{T+2}$ is binarized into a binary mask by Otsu. $\beta_{1}$, $\beta_{2}$, $\gamma _{1}$, $\gamma _{2}$, $\delta _{1}$, $\delta _{2}$ are parameters selected by considering the resolution of aerial images and observing the the datasets. 

If $f^{T+2}$ meets all the requirements listed above, we consider it as a reasonable output, and can be reused as the label in the next epoch of further training. Otherwise, there are three choices: If $f^{T+2}$ fails to meet the first requirement, the initial label will be used for the next round of training; if $f^{T+2}$ fails to meet the second requirement, the label for next round will be the initial label; if $f^{T+2}$ are dramatically different from the previous $f^{T}$, but meet the requirements one and two, the label will not be updated. 

With a reasonable output, morphology operations are further utilized for label refinement. Specifically, the output $f^{T+2}$ is first transformed into a binary mask by the Otsu thresholding:
\begin{equation}
Mask = OTSU(f^{T+2}).
\end{equation}

To minimize the impact of the background interference including shadow, an opening operation with $5\times 5$ structuring element $SE_{1}$ is applied to the produced $Mask$. This can help reduce the false positives in the mapping. Then, another dilation with $3\times 3$ structuring element $SE_{2}$ is conducted to improve the completeness of the regions:
\begin{equation}
{Mask}'=Mask\circ SE_{1} = (Mask \oplus SE_{1} )\ominus SE_{1},
\end{equation}
\begin{equation}
GT^{T+2}={Mask}' \oplus SE_{2}, 
\end{equation}
where $\circ$, $\oplus$, $\ominus$ denotes the opening, dilation and erosion operation, respectively.

After the new labels are updated each time, the mapping network is further trained. 
\begin{figure*}[htp]
    \centering
    \includegraphics[width=18cm]{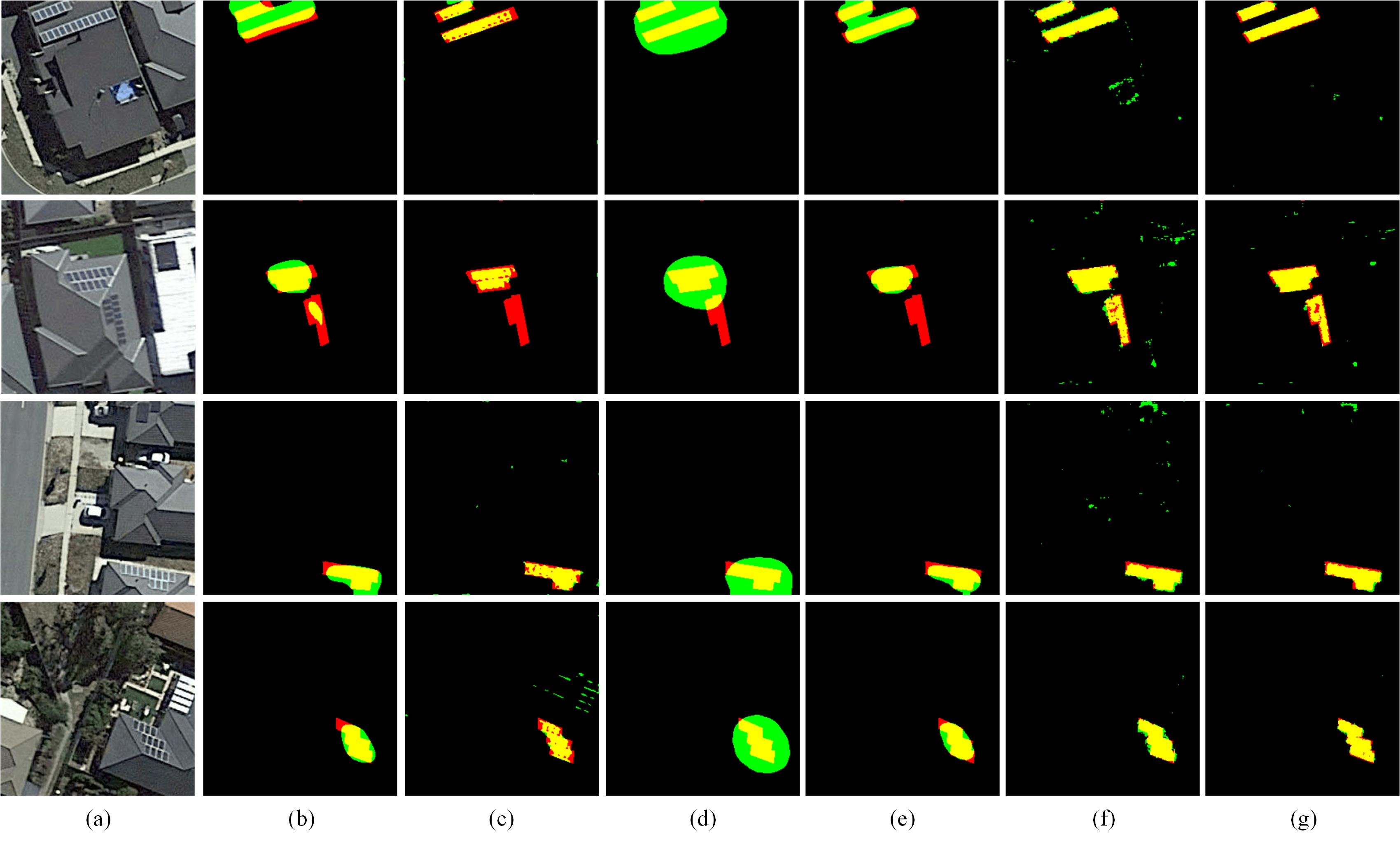}
    \caption{Solar panel mapping results marked with true positives (Yellow), true negatives (black), false positives (Green) and false negatives (Red). (a) Original images. (b) DeepSolar. (c) HWSL. (d) GradCAM$_{5}$. (e) GradCAM$_{4}$. (f) PS-CNN. (g) PS-CNNLC.}
    \label{fig:my_label}
\end{figure*}

\begin{table}[htbp]
\centering 
\caption{Quantitative Performance Comparison on The Aerial Dataset}
\begin{center}
\begin{tabularx}{0.45\textwidth}{l c c c c c}
\toprule[1.3pt]
\textbf{Method} & \textbf{AC} &\textbf{AUC} & \textbf{P}& \textbf{R}& \textbf{$F_{\theta}$} \\[1.1ex] 
\toprule[1.3pt]
DeepSolar& 0.9776&	0.9841&	0.6767&	0.7793&	0.6908 \\[1.1ex] 
\toprule[0.5pt]
HWSL&0.9772&	0.8802&	0.7910&	0.5741&	0.6906
\\[1.1ex] 
\toprule[0.5pt]
GradCAM$_{4}$&0.9853&	\textbf{0.9944}&	0.8195&	0.7811&	0.7999
\\[1.1ex] 
\toprule[0.5pt]
GradCAM$_{5}$&0.9374&	0.9774&	0.3530&	\textbf{0.9079}&	0.4079
\\[1.1ex]
\toprule[0.5pt]
PS-CNN&0.9896&	0.9941&	0.8329&	0.8938&	0.8419
\\[1.1ex]
\toprule[0.5pt]
PS-CNNLC&\textbf{0.9912}&	0.9929&	\textbf{0.9263}&	0.8245&	\textbf{0.8975}
\\[1.1ex]
\toprule[1pt]
\end{tabularx}
\label{tab1}
\end{center}
\end{table}
\section{Experiments}
\begin{figure}
    \centering
    \includegraphics[width=9cm]{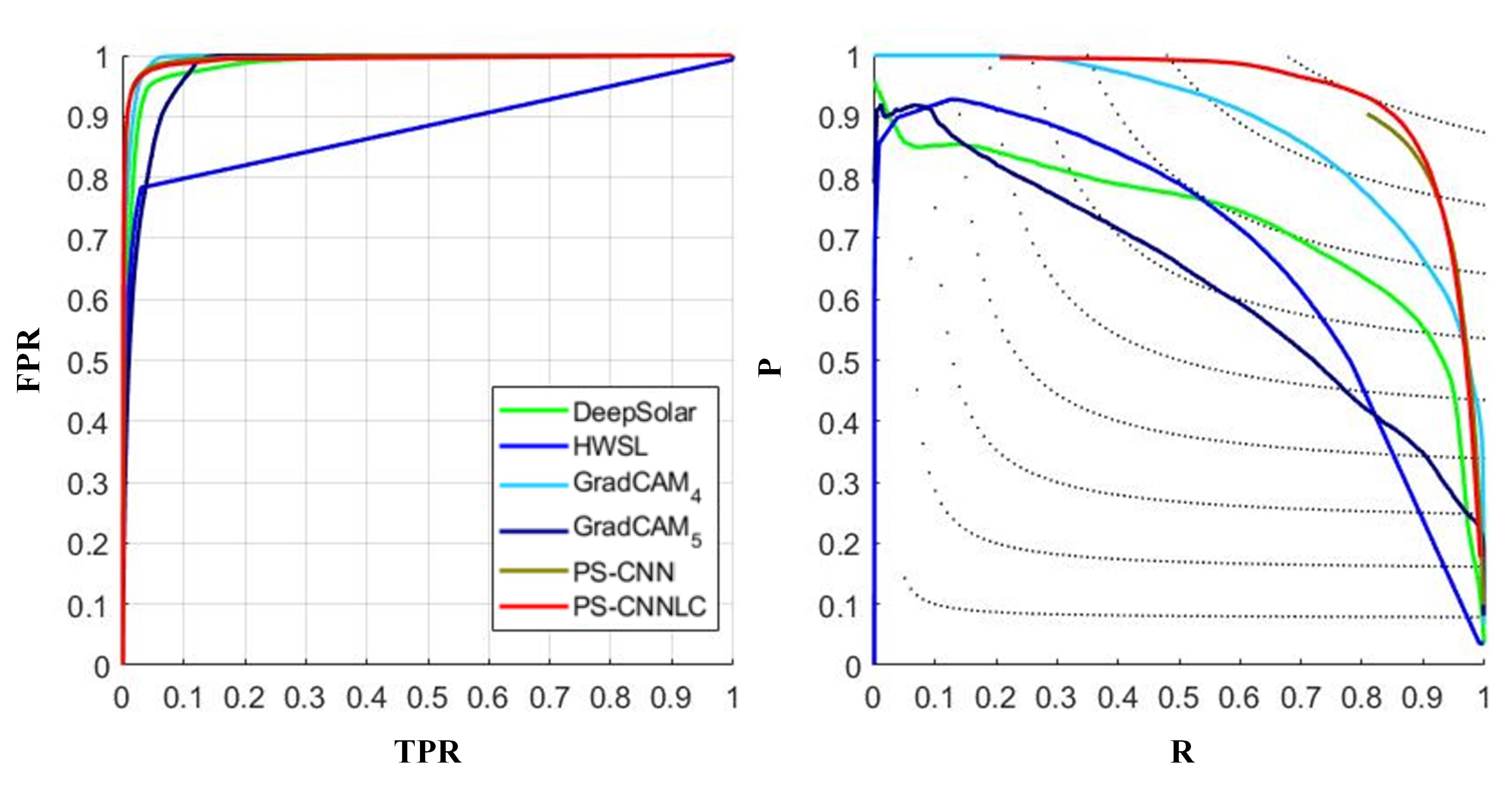}
    \caption{ROC curves and PR curves of different mapping methods.}
    \label{fig:my_label}
\end{figure}

\subsection{Experimental Settings}
\subsubsection{Dataset}
The proposed method was evaluated on an aerial image (RGB) dataset based on the Google Static Map API. All the images were collected in Canberra, ACT, Australia in April 2019 with the spatial resolution higher than 0.3 meters. For the image-level classification network, the training set contain 428 positive samples and 446 negative samples while the testing set comprises 25 positive and 25 negative samples. Manually annotated image-level labels were utilized for training of the classification network. Beside the 428 positive samples, 2000 unlabeled samples are added to produce more positive samples and corresponding pseudo labels for the target mapping network. As a result, 1493 training samples were available for the development of the fully supervised mapping network. The testing set for the target mapping network contains 25 samples. All samples employed have the size of $256\times 256 \times 3$.
\subsubsection{Implementation Details}
The proposed method was implemented with Tensorflow and trained on a single NVIDIA Tesla P4 GPU. The weights of all the convolutional layers in the classification network was initialized by the VGG16 pre-trained over the ImageNet. The fully connected layers were randomly initialized with a normal distribution. RMSprop optimizer with learning rate $10^{-4}$ was used to train the classification network. The batch size is 16. The fully supervised mapping network was trained by the RMSprop optimizer with initial learning rate $10^{-3}$ and batch size 15. The learning rate was decreased by a factor of 0.8 every 20 epochs. The mapping network was firstly trained 800 epochs without the label correction strategy and then 40 epochs with the label correction strategy. In the label correction strategy, we set $\beta _{1}=0.01$, $\beta  _{2}=0.1$, $\gamma _{1}=0.6$, $\gamma _{2}=1.4$, $\delta _{1}=0.8$, $\delta _{2}=1.2$.  
\subsubsection{Evaluation Metrics}
For quantitative evaluation of the mapping results, we used overall accuracy (AC), receiver operator characteristic (ROC) curve, area under the curve (AUC), Precision, Recall and F-measure. With a generated map, varying thresholds are applied to binarized the map. For every threshold utilized, the false positive rate and true positive rate are calculated for ROC curve; the overall accuracy, Precision $P$, Recall $R$ and F-measure are defined as follows:
\begin{equation}
AC = \frac{TP+TN}{TP+TN+FP+FN},
\end{equation}
\begin{equation}
P = \frac{TP}{TP+FP},
\end{equation}
\begin{equation}
R = \frac{TP}{TP+FN},
\end{equation}
\begin{equation}
    F_{\theta} = \frac{(1+\theta^{2})\cdot P\cdot R}{\theta^{2}\cdot P + R}, 
\end{equation}
where $TP$, $TN$, $FP$, $FN$ denotes the total number of true positive, true negative, false positives and false negative pixels in the binary segmentation map. In this paper, $\theta^{2} =0.3$.

Higher overall accuracy, Precision, Recall, F-measure and AUC indicate better performance. 
\subsection{Comparisons with State-of-the-Art Methods}
\begin{figure*}[htp]
    \centering
    \includegraphics[width=18cm]{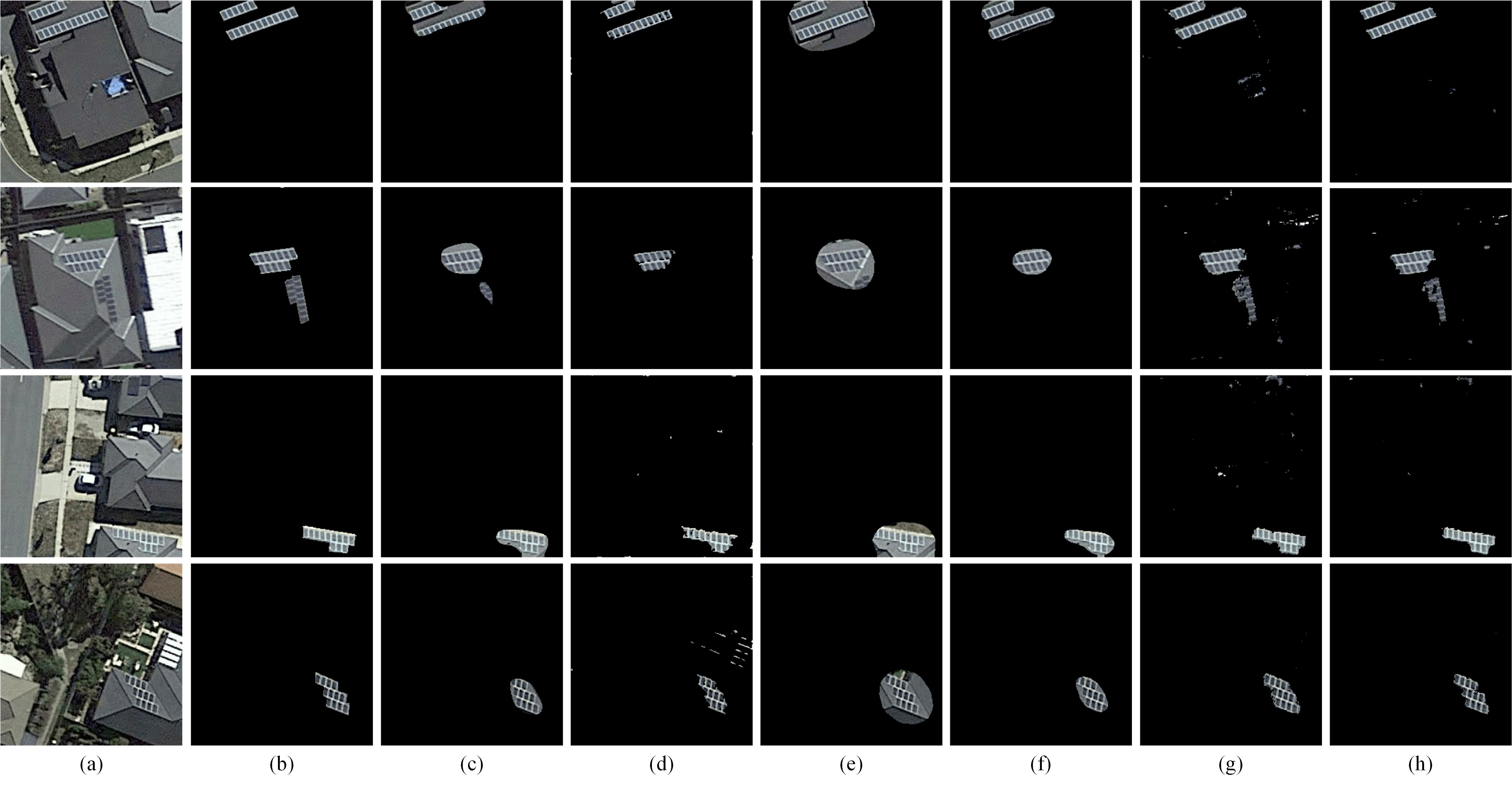}
    \caption{Extraction results by different methods. (a) Original images. (b) Ground Truth. (c) DeepSolar. (d) HWSL. (e) GradCAM$_{5}$. (f) GradCAM$_{4}$. (g) PS-CNN. (h) PS-CNNLC.}
    \label{fig:my_label}
\end{figure*}
\subsubsection{Comparison Methods}
We compared the proposed method with two state-of-the-art methods including DeepSolar, a weakly supervised CNN-based method proposed for solar panel mapping and Hierarchical Weakly Supervised Learning (HWSL), a weakly-supervised CNN-based method  proposed for residential area semantic segmentation in remote sensing images. We used the source codes of DeepSolar and HWSL released by the authors. All the parameters were set according to the authors' descriptions in the literature.
\subsubsection{Ablation Study}
In order to further reveal the validity of the proposed pseudo supervised method, ablation study was also conduct. We investigated the influence of the employment of different convolutional layers in activation map generation, and the impact of label correction strategy. The four cases for the ablation study are:
\begin{itemize}
\item GradCAM$_{5}$: mapping results using GradCAM with $Conv5\_3$ in classification network.
\item GradCAM$_{4}$: mapping results using GradCAM with $Conv4\_3$ in classification network.
\item PS-CNN: the proposed the pseudo supervised convolutional neural network trained with only initial pseudo labels generated by GradCAM with $Conv4\_3$ in classification network.
\item PS-CNNLC: the proposed pseudo supervised solar panel mapping method with label correction strategy. Initial pseudo labels were generated by GradCAM with $Conv4\_3$ in classification network.
\end{itemize}

\subsubsection{Discussion}
We evaluated of the state-of-the-art methods and ablation study visually and quantitatively. The solar panel mapping results by different methods are showed in Figure 4. In Figure 5, results are marked with true positives (Yellow), true negatives (black), false positives (Green) and false negatives (Red). The quantitative performance is summarized in Table 1, with highest values of every metric presented in bold. The ROC curve and PR curve are shown in Figure 6. Figure 7 shows the extraction results.

The results reveal that the proposed method substantially outperforms other weakly supervised methods including DeepSolar and HWSL. Deepsolar provides rough localization for solar panels with certain false negatives and false positives. HWSL shows better boundary maintenance, but the extracted regions are incomplete. In addition, some special ground objects similar to solar panels are also extracted by mistake. Compared with Deepsolar and HWSL, our proposed network provides well-localized solar panel identification and well-preserved boundary completeness, with lower false alarm rate. 

In the ablation study, GradCAM$_{4}$ is the baseline of our work. From Figure 4(e), we can see that compared with the results of GradCAM$_{5}$, the results of GradCAM$_{4}$ are more accurate. The proposed PS-CNN combines the advantages of weakly and fully supervised learning, and outperforms other weakly supervised methods including DeepSolar and HWSL both visually and quantitatively. By comparing the results of PS-CNN and PS-CNNLC, the impact of the label correction strategy can be revealed: it helps reduce the false positive to a low extent and provide well-defined boundaries. The quantitative results in Table 1 show that PS-CNN and PS-CNNLC have higher mapping accuracy and better overall performance in precision and recall.      

\section{Conclusion}
In this paper, we proposed a pseudo supervised solar panel mapping method based on deep convolutional networks with label correction strategy in aerial images. In the proposed method, a classification network supervised by image-wise labels is trained to identify images with solar panels. Pseudo labels are produced with activation maps generated by the classification network and GradCAM. Then, a fully supervised mapping network is constructed in an encoder and decoder manner to produce full-resolution detection for solar panels. In order to improve the accuracy, a label correction strategy is developed to refine the labels in the training phase. Accuracy improvement is achieved when the mapping network is trained with progressively corrected labels. 

\bibliographystyle{IEEEtran}
\bibliography{IEEEabrv,mybibfile}

\end{document}